\documentstyle[12pt,titlepage]{article} 
\newcommand{\ba}{\begin{array}}
\newcommand{\ea}{\end{array}}
\newcommand{\bd}{\begin{displaymath}}
\newcommand{\ed}{\end{displaymath}}
\newcommand{\be}{\begin{equation}}
\newcommand{\ee}{\end{equation}}
\newcommand{\bea}{\begin{eqnarray}}
\newcommand{\eea}{\end{eqnarray}}

\def\bra{\langle}
\def\ket{\rangle}

\def\b{\beta}
\def\g{\gamma}

\def\e{\epsilon}

\def\l{\lambda}
\def\m{\mu}
\def\n{\nu}
\def\t{\tau}

\def\L{\Lambda}

\begin{document}
\begin{titlepage}

\begin{flushright}
\begin{tabular}{l}
MRI-PHY/96-37\\
December, 1996\\
hep-ph/9612447
\end{tabular}
\end{flushright}
\vskip .6cm

\begin{center}
{\Large\bf SOME IMPLICATIONS OF A SUPERSYMMETRIC MODEL WITH R-PARITY
BREAKING BILINEAR INTERACTIONS}\\
{\large  Sourov Roy$^1$ and Biswarup Mukhopadhyaya$^2$\\}
Mehta Research Institute,
10 Kasturba Gandhi Marg,
Allahabad - 211 002, INDIA \\
\end{center}
\vskip .5cm

\begin{center}
{\bf ABSTRACT}\\
\end{center}

\noindent We investigate a supersymmetric scenario where R-parity is 
explicitly broken through a term bilinear in the lepton and Higgs superfields 
in the superpotential. We show that keeping such a term alone can lead to
trilinear interactions, similar to those that are parametrized by
$\lambda$-and ${\lambda}'$ in the literature, involving the physical fields.
The upper limits of such interactions are predictable from the constraints
on the parameter space imposed by the lepton masses and the neutrino mass 
limits. It is observed that thus the resulting trilinear interactions are
restricted to values that are smaller than the existing bounds on
most of the $\lambda$-and ${\lambda}'$-parameters. Some phenomenological
consequences of such a scenario are discussed.  

\hspace*{\fill}
\vskip .2in

\noindent
PACS NOS. : 12.60.Jv, 13.10.+q, 14.80.Ly

\hspace*{\fill}
\vskip .5in

\noindent
$^{1}$E-mail : ~~~sourov@mri.ernet.in \\
$^{2}$E-mail : biswarup@mri.ernet.in

\end{titlepage}

\textheight=8.9in

\section{INTRODUCTION}

 ~~ It is being increasingly realised by those engaged in the search for
supersymmetry (SUSY) \cite{rev} that the principle of R-parity conservation, 
assumed to be sacrosanct in the prevalent search strategies, is not in 
practice inviolable. The R-parity of a particle is defined as 
$R = (-1)^{L + 3B + 2S}$, and can be violated if either baryon (B) or
lepton (L) number is not conserved in nature, a fact perfectly compatible
with the non-observation of proton decay. This is because, whereas the 
violation of B or L, taken {\it singly}, is inadmissible in the standard model 
(SM) where all the elementary baryons and leptons are fermions, the SUSY 
version of the SM allows it by virtue of the scalar quarks and leptons that are
part of the particle spectrum.  

 Under R-parity violation the phenomenology changes considerably \cite{rp},
the most important consequence being that the lightest supersymmetric 
particle (LSP) can decay now. However, the way in which R-parity can be 
violated is not unique; different types of R-violating interaction terms 
can be written down, leading to different observable predictions. In addition, 
R-parity can be violated spontaneously, in stead of explicitly, whence another
class of interesting effects are expected \cite{spn}. If the phenomenology
of R-parity breaking has to be understood, and the consequent modifications 
in the current search strategies have to be effectively implemented, then it 
is quite important to explore the full implication of each possible R-breaking
scheme. In this paper we probe some aspects of one such scheme, 
namely, where lepton number violation has its origin in terms bilinear in the 
lepton and Higgs superfields in the superpotential \cite{daw}.

  The R-conserving part of the minimal supersymmetric standard model
(MSSM) is of the following form in terms of superfields:

\begin{equation}
W_{MSSM} = \epsilon_{ab}[{\mu} H_1^a H_2^b + h_{ij}^l L_i^a H_1^b E_j^c +  
h_{ij}^d Q_i^a H_1^b D_j^c + h_{ij}^u Q_i^a H_2^b U_j^c]
\end{equation}

\noindent where $(a,b)$ are SU(2) indices, $(i,j)$ are generation indices
and the superscript $c$ denotes right-handed chiral superfields. Here
$Q = {u \choose d}$, $L = {{{\n}_{l}} \choose l}$ and ${H_1}$, ${H_2}$ are
the Higgs superfields that gives masses to the down- and up-type quark 
superfields. If now R-breaking interactions are incorporated, 
the superpotential takes the form \cite{bhg}

\begin{equation}
W = W_{MSSM} + W_L + W_B
\end{equation}

\noindent 
with

\begin{equation}
W_L = \epsilon_{ab}[\lambda_{ijk} L_i^a L_j^b E_k^c + 
\lambda_{ijk}' L_i^a Q_j^b D_k^c + \epsilon_i L_i^a H_2^b]
\end{equation}

\noindent 
and

\begin{equation}
W_B = \lambda_{ijk}'' U_i^c D_j^c D_k^c
\end{equation}

 Obviously, both $W_L$ and $W_B$ cannot be present if the proton has to be 
stable. In the rest of this paper we shall concentrate on the case where
only lepton number is violated.

$W_B$ as well as the first two terms in $W_L$ have received a lot of attention
in recent times, and constraints have been derived on them from existing 
experimental data \cite{limit}. However, the term $\epsilon_i L_i^a H_2^b$  
is also a viable agent for R-parity breaking. It is particularly interesting 
for the fact that it can trigger a mixing between charginos and charged 
leptons as well as between neutralinos and neutrinos, resulting in 
observable effects that are not to be seen with the $\lambda$-and 
$\lambda'$-terms alone. One of these distinctive effects is that, the 
lightest neutralino can decay invisibly into three neutrinos, which is not 
possible if only the first two terms in $W_L$ are present. Some other 
implications, especially those in the scalar sector of the theory, have 
been investigated recently in the literature \cite{anj}. The significance of 
such bilinear R-violating interactions is further emphasized by the following 
observations:

\noindent(1) Although it may seem possible to rotate away the $LH_2$-terms
by redefining the lepton and Higgs superfields, their effect is bound to
show up via the scalar potential \cite{anj}.

\noindent(2) Even if one may rotate away these terms at one enrgy scale,
they reappear at another as the couplings evolve radiatively \cite{bp}.

\noindent(3) the $\lambda$-and $\lambda'$-terms themselves give rise to
the bilinear terms at the one-loop level \cite{carlos}.

\noindent(4) It has been argued that if one wants to subsume R-parity
violation in a Grand Unified Theory (GUT), then the trilinear interactions
in $W_L$ naturally come out to be rather small in magnitude 
($O(10^{-3})$ or so) \cite{hall}. However, the superrenormalizable bilinear 
terms are not subjected to such requirements {\it a priori}.

We perform an analysis here keeping ${\epsilon}LH_2$ 
as the {\it only} R-parity violating term in the theory \cite{addl}.
Moreover, for reasons that we shall discuss below, we are incorporating 
such a term only for the third generation lepton superfield $L_3$. 
We shall see that after one incorporates the effect of mixing, such 
a term can give rise to trilinear interactions among the physical states,
which are very similar in nature to those induced by the $\lambda$'s and 
the $\lambda'$'s. All these interactions are derived in section 2, 
together with the gauge boson couplings of the
lepton-chargino and neutrino-neutralino physical states. It is 
interesting to note that the parameters giving rise to these interactions
are constrained by the $\tau$-and $\nu_\tau$ masses. Thus it is possible
to predict the maximum possible values for the couplings for any given
set of parameters of the MSSM.
In section 3  we discuss these  constraints  and some of their phenomenological
consequences. Our conclusions are summarised in section 4. The detailed 
forms of some formulas of section 2 are presented in the appendix.

\section{The Formalism}

\noindent As has been stated before, we consider a superpotential of the form
\bea
W = W_{MSSM} + {\e}{L_3}{H_2}
\eea
where the $\rm SU(2)$ indices have been suppressed. The simplification achieved
by letting only the third generation mix with the Higgs superfield can be
justified if one notes that the value of ${\e}_i$ for a particular 
generation is constrained severely by the upper limit on the neutrino mass in 
that generation. Since the $\t$-neutrino mass has the least restrictive 
laboratory bound of 24 MeV \cite{pdg}, only ${\e}_3$ (to be called $\e$ 
hereafter) can be large enough to be phenomenologically significant.

An immediate consequence of a non-zero $\e$ is the mixing between the
charged leptons and the charginos as well as between neutrinos and neutralinos.
The other quantity that can trigger such mixing is a non-zero vacuum 
expectation value (vev) of ${\tilde \n}_{\t}$. This vev leads
to off-diagonal ${\n}_{\t}-{\tilde Z}$ and ${\t}-{\tilde W}$ terms in the 
current eigenstate basis \cite{ross}.

In such a situation, the $(3{\times}3)$ chargino mass matrix is 
\bea
M_{{\tilde \chi}^{\pm}} = \left(\matrix{M & -g {v_2} & 0\cr
  -g {v_1} & \mu & f {v_3}\cr
  -g {v_3} & \e & -f{v_1}\cr}\right)
\eea  
where $v_1 = \bra{H_1}\ket, ~~v_2 = \bra{H_2}\ket, 
~~v_3 = \bra{{\tilde \n}_{\t}}\ket$ and
~$f = h^{l}_{33} = {\frac {m_{\t}}{v_1}}$,
~$M$ being the $\rm SU(2)$ gaugino mass parameter. Here we have assigned
$(-i{\bar {{\tilde W}^-}}, \bar {{{\tilde H}_1}^-}, {\bar {{\t}_L}^-})$ 
along the rows and 
$(-i{\bar {{\tilde W}^+}}, \bar {{{\tilde H}_2}^+}, {\bar {{\t}_R}^+})$
along the columns. Similarly, the extended neutralino mass matrix in the
basis $(-i{ {\tilde A}}, -i{\tilde Z}, {{\tilde H}_1}^0, {{\tilde H}_2}^0,
{\n}_{\t})$ is given by 
\bea
M_{{\tilde \chi}^0} = \left(\matrix{M_{\tilde A} & \frac1 2(M_{\tilde Z} 
- M_{\tilde A}){\rm tan}
  2 {\theta_W} & 0 & 0 & 0\cr
  \frac1 2(M_{\tilde Z} - M_{\tilde A}){\rm tan}
  2 {\theta_W} & M_{\tilde Z} & -\frac{g {v_1}}
  {\sqrt{2} {\rm cos} {\theta_W}} & 
  \frac{g {v_2}}
  {\sqrt{2} {\rm cos} {\theta_W}} & 
  -\frac{g {v_3}}
  {\sqrt{2} {\rm cos} {\theta_W}}\cr
  0 & -\frac{g {v_1}}
  {\sqrt{2} {\rm cos} {\theta_W}} & 
  0 & -{\m} & 0\cr
  0 & \frac{g {v_2}}
  {\sqrt{2} {\rm cos} {\theta_W}} & -{\m} & 0 & -{\e}\cr
  0 & -\frac{g {v_3}}
  {\sqrt{2} {\rm cos} {\theta_W}} & 0 & -{\e} & 0\cr}\right)
\eea  
  with 
\bea
M_{\tilde A} = M'{\rm cos}^2 {\theta}_W + M{\rm sin}^2 {\theta}_W
\eea
\bea
M_{\tilde Z} = M'{\rm sin}^2 {\theta}_W + M{\rm cos}^2 {\theta}_W
\eea  
$M'$ and $M$ being respectively the $U(1)$ and $SU(2)$ gaugino mass parameters.

The diagonalisation of 
$M_{{\tilde \chi}^{\pm}}$ and $M_{{\tilde \chi}^0}$
is straightforward; one thus obtains two $(3{\times}3)$ matrices $\rm U$
and $\rm V$ for the right-handed and left-handed chargino respectively and
a $(5{\times}5)$ mixing matrix $\rm N$ for the neutralinos. These
correspond to their $\rm {MSSM}$ forms in the proper limit.   

\noindent Let us now consider the scalar sector in this scenario. The
scalar potential, including the third generation sleptons, is given by
\bea
V~~~ =~~~~~~~~~~ m_{1}^2 {H_{1}^{\dagger}}{H_1} + m_{2}^2 {H_{2}^{\dagger}}
{H_2} + m_{\tilde L}^2 {{{\tilde {\t}}_L}^{\dagger}}{{\tilde {\t}}_L} + 
m_{\tilde R}^2 {{{\tilde {\t}}_R}^{\dagger}}{{\tilde {\t}}_R}
+ {m_{\tilde {{\n}_{\t}}}^2} {{{\tilde {\n}}_{\t}}^{\dagger}}
{{\tilde {\n}}_{\t}}\cr + 
{f^2} {H_{1}^{\dagger}}{H_1}({{\tilde L}^{\dagger}}{\tilde L} + 
{{{\tilde {\t}}_R}^{\dagger}}{{\tilde {\t}}_R}) + 
{f^2}{{\tilde L}^{\dagger}}{\tilde L}
{{{\tilde {\t}}_R}^{\dagger}}{{\tilde {\t}}_R}
+ {\m} f[{H_{2}^{\dagger}}{\tilde L}{{{\tilde {\t}}_R}^{\dagger}}
 + {{\tilde L}^{\dagger}} {H_2}
{{\tilde {\t}}_R}]\cr - {\e} f [{H_{1}^{\dagger}}{{\tilde {\t}}_R}
{H_2} + {{H_1}{{{\tilde {\t}}_R}^{\dagger}}{H_{2}^{\dagger}}] + 
{\m}{\e}[{\tilde L}{H_{1}^{\dagger}} + {{\tilde L}^{\dagger}}{H_1}] 
- {f^2}{H_{1}^{\dagger}}{\tilde L}
({H_{1}^{\dagger}}{\tilde L})^{\dagger}}\cr 
+ {B_1}{\m}{({{\phi}_{1}^0}
{{\phi}_{2}^0} - {{\phi}_{1}^-}{{\phi}_{2}^+} + {{{\phi}_{2}^0}^{\dagger}}
{{{\phi}_{1}^0}^{\dagger}} - {{{\phi}_{2}^+}^{\dagger}}
{{{\phi}_{1}^-}^{\dagger}})} + A f ({{\tilde {\t}}_L} {{\phi}_{1}^0} -
{{\tilde {\n}}_{\t}}{{\phi}_{1}^-}){{{\tilde {\t}}_R}^{\dagger}}\cr 
+{B_2}{\e}{({{\tilde {\n}}_{\t}}
{{\phi}_{2}^0} - {{\tilde {\t}}_L}{{\phi}_{2}^+} + {{{\phi}_{2}^0}^{\dagger}}
{{\tilde {\n}}_{\t}}^{\dagger} - {{{\phi}_{2}^+}^{\dagger}}
{{\tilde {\t}}_L}^{\dagger}})  
+ A f ({{{\tilde {\t}}_L}^{\dagger}} {{{\phi}_{1}^0}^{\dagger}} -
{{{\tilde {\n}}_{\t}}^{\dagger}}{{{\phi}_{1}^-}^{\dagger}})
{{{\tilde {\t}}_R}}\cr + {\frac 1 8} (g^2 +{g'}^2) 
[{({H_{1}^{\dagger}}{H_1} - {H_{2}^{\dagger}}{H_2})}^2] + {\frac 1 2} {g^2}
{\vert {H_{1}^{\dagger}}{H_2} \vert}^2\cr - {\frac 1 2}{{g'}^2} 
{{{\tilde {\t}}_R}^{\dagger}}{{\tilde {\t}}_R}
{({{\tilde L}^{\dagger}}{\tilde L} + {H_{1}^{\dagger}}{H_1} 
- {H_{2}^{\dagger}}{H_2})}\cr + {\frac 1 4} {g^2} 
{({{\tilde {\n}}_{\t}}^{\dagger}{{\tilde {\t}}_L}{{{\tilde {\t}}_L}^{\dagger}}
{\tilde {{\n}_{\t}}})} + {\frac 1 2} {g^2} {({{\tilde {\n}}_{\t}}^
{\dagger}{{\tilde {\t}}_L} {{{\phi}_{1}^-}^{\dagger}}{{\phi}_{1}^0} 
+{{{\tilde {\t}}_L}^{\dagger}}{{\tilde {\n}}_{\t}}} 
{{{\phi}_{1}^0}^{\dagger}}{{\phi}_{1}^-})\cr
+ {\frac 1 4} {g^2} {({{\tilde {\n}}_{\t}}^
{\dagger}{{\tilde {\n}}_{\t}} {{{\phi}_{1}^0}^{\dagger}}{{\phi}_{1}^0} 
+{{{\tilde {\t}}_L}^{\dagger}}{{\tilde {\t}}_L}{{{\phi}_{1}^-}^{\dagger}} 
{{\phi}_{1}^-} - {{\tilde {\n}}_{\t}}^{\dagger}
{{\tilde {\n}}_{\t}}{{{\phi}_{1}^-}^{\dagger}}{{\phi}_{1}^-}
- {{{\tilde {\t}}_L}^{\dagger}}{{\tilde {\t}}_L}{{{\phi}_{1}^0}^{\dagger}}
{{\phi}_{1}^0})}\cr + {\frac 1 4} {g^2} {({{\tilde {\n}}_{\t}}^
{\dagger}{{\tilde {\n}}_{\t}} {{{\phi}_{2}^+}^{\dagger}}{{\phi}_{2}^+} 
+{{{\tilde {\t}}_L}^{\dagger}}{{\tilde {\t}}_L}{{{\phi}_{2}^0}^{\dagger}} 
{{\phi}_{2}^0} - {{\tilde {\n}}_{\t}}^{\dagger}
{{\tilde {\n}}_{\t}}{{{\phi}_{2}^0}^{\dagger}}{{\phi}_{2}^0}
- {{{\tilde {\t}}_L}^{\dagger}}{{\tilde {\t}}_L}{{{\phi}_{2}^+}^{\dagger}}
{{\phi}_{2}^+})}\cr + {\frac 1 2} {g^2} {({{\tilde {\n}}_{\t}}^
{\dagger}{{\tilde {\t}}_L} {{{\phi}_{2}^0}^{\dagger}}{{\phi}_{2}^+} 
+{{{\tilde {\t}}_L}^{\dagger}}{{\tilde {\n}}_{\t}}} 
{{{\phi}_{2}^+}^{\dagger}}{{\phi}_{2}^0})}+ {\frac 1 4} {{g'}^2} 
{{{\tilde {\t}}_L}^{\dagger}}{{\tilde {\t}}_L}
{({H_{1}^{\dagger}}{H_1} - {H_{2}^{\dagger}}{H_2})}
+ {\frac 1 4} {{g'}^2}{({{\tilde {\n}}_{\t}}^{\dagger}{{\tilde {\t}}_L}
{{{\tilde {\t}}_L}^{\dagger}} {\tilde {{\n}_{\t}}})\cr
+ {\frac 1 8} {({g'}^2 + g^2)}{\{({{{\tilde {\n}}_{\t}}^{\dagger}
{\tilde {\n}}_{\t})}^2 + {({{\tilde {\t}}_L}^{\dagger}{\tilde {\t}}_L)}^2\}}
\eea
where $\tilde L = {{\tilde {\n}}_{\t} \choose {\tilde {\t}}}_L$,
${H_1} = {{\phi}_{1}^0 \choose {\phi}_{1}^-}$,
${H_2} = {{\phi}_{2}^+ \choose {\phi}_{2}^0}$. $A$ is the SUSY breaking trilinear
soft term and ${B_1}$, ${B_2}$ are the bilinear soft terms.

The mass-squared matrices for the neutral
scalars, neutral psedoscalars and charged scalars are given respectively by
 
\bea
M_{s}^2 = \left(\matrix
{m_{1}^2+2{\l}c+4{\l}v_{1}^2 & -4{\l}{v_1}{v_2}+{B_1}{\m} &
4{\l}{v_1}{v_3}+{\m}{\e}\cr
-4{\l}{v_1}{v_2}+{B_1}{\m} & m_{2}^2-2{\l}c+4{\l}v_{2}^2 & 
-4{\l}{v_3}{v_2}+{B_2}{\e}\cr 
4{\l}{v_1}{v_3}+{\m}{\e} & 
 -4{\l}{v_3}{v_2}+{B_2}{\e} & {m_{{\tilde {\n}}_{\t}}^2}
 +2{\l}c+4{\l}v_{3}^2\cr}
 \right)
\eea
\bea 
M_{p}^2 = \left(\matrix
{m_{1}^2+2{\l}c & -{B_1}{\m} &
{\m}{\e}\cr
-{B_1}{\m} & m_{2}^2-2{\l}c & 
-{B_2}{\e}\cr 
{\m}{\e} & 
-{B_2}{\e} & {m_{{\tilde {\n}}_{\t}}^2}+2{\l}c\cr}
 \right)
\eea
and
\bea 
{M_c}^2 = \left(\matrix
{r - {\frac{1}{4}}{g'}^2 c & 
 -{B_1}{\m} + {\frac{1}{2}}g^2 {v_1} {v_2} & -{B_2}{\e} + {\frac{1}{2}}
 g^2 {v_2} {v_3} & -{\e} f {v_1}\cr
 -{B_1}{\m} + {\frac{1}{2}}g^2 {v_1} {v_2} & 
 s + {\frac{1}{4}}{g'}^2 c &
 {\m}{\e} + {\frac{1}{2}}g^2 {v_1}{v_3} & -{\e} f {v_2} + A f {v_3}\cr
 -{B_2}{\e} + {\frac{1}{2}} g^2 {v_2} {v_3} & 
  {\m}{\e} + {\frac{1}{2}}g^2 {v_1}{v_3} & 
 p + {\frac{1}{4}} g^2 t + {\frac{1}{4}}{g'}^2 c  &  
 {\m} f {v_2}-A f {v_1}\cr
  -{\e} f {v_1} & -{\e} f {v_2} + A f {v_3} & {\m} f {v_2}-A f {v_1} & 
  q -{\frac{1}{2}} {g'}^2 c + f^2 {v_3}^2\cr}  
  \right)
\eea  
with  

$$r = m_{2}^2 + {\frac 1 4}{g^2}(v_{1}^2 + v_{2}^2 + v_{3}^2)$$
$$s = m_{1}^2 + {\frac 1 4}{g^2}(v_{1}^2 + v_{2}^2 - v_{3}^2)$$
$$p = m_{\tilde L}^2 + f^2 v_{1}^2$$
$$q = m_{\tilde R}^2 + f^2 v_{1}^2$$
$$t = (-v_{1}^2 + v_{2}^2 + v_{3}^2)$$
$$c = (v_{1}^2 -v_{2}^2 + v_{3}^2)$$
$$\l = (g^2 + {g'}^2)/8$$
Here the real and imaginary part of ${\tilde {\n}}_{\t}$ enter into 
${M_s}^2$ and ${M_p}^2$ respectively.
The corresponding diagonalising matrices that control the mixing in those
sectors are described here as $S$, $P$ and $C$.

The scalar sector is subject to the following constraints \cite{simm} :

\noindent(1) The extremization of the neutral part of the potential leads to
\bea
(m_{1}^2 + 2 {\l} c) {v_1} + {B_1} {\m} {v_2} + {\m}{\e}{v_3} = 0
\eea
\bea 
(m_{2}^2 - 2 {\l} c) {v_2} + {B_1} {\m} {v_1} + {B_2}{\e}{v_3} = 0
\eea
\bea 
({m_{{\tilde {\n}}_{\t}}^2} + 2 {\l} c) {v_1} 
+ {B_1} {\m} {v_2} + {\m}{\e}{v_3} = 0 
\eea
Furthermore, the second derivatives with respect to the neutral fields at
the extremum must be all positive.

\noindent(2) The potential must be bounded from below \cite{stab}. 
The resulting condition is 
\bea
m_{1}^2 (v_{2}^2 - v_{3}^2) + m_{2}^2 v_{2}^2 
+ {m_{{\tilde {\n}}_{\t}}^2} v_{3}^2 + 2 {\m} {\e} {(v_{2}^2 
- v_{3}^2)}^{\frac 1 2}
{v_3}\cr + 2 {B_1} {\m} {(v_{2}^2 - v_{3}^2)}^{\frac1 2}{v_2} 
+ 2 {B_2} {\e}{v_2}{v_3}~~{\ge}~~0
\eea  
Note that setting $v_3 = 0$ above give us the corresponding condition in MSSM.

\noindent(3) Gauge symmetry breaking requires that the minimum of the potential
has to be negative \cite{gunn}. This implies 
\bea
X_{min}~~\le~~0
\eea
where $X_{min}$ is the lowest eigenvalue of the matrix
\begin{eqnarray}
\left(\matrix
{m_{1}^2 & {B_1}{\m} &
{\m}{\e}\cr
{B_1}{\m} & m_{2}^2 & 
{B_2}{\e}\cr 
{\m}{\e} & 
{B_2}{\e} & {m_{{\tilde {\n}}_{\t}}^2}\cr}
 \right)
\end{eqnarray} 

\noindent(4) All the eigenvalues of the $M_{s}^2, M_{p}^2$ and $M_{c}^2$ have 
to be non-negative. This leads to the necessary (but not sufficient) conditions
that $B_1$ and $\m$, as also $B_2$ and $\e$, are of opposite signs.

Once the five mass matrices mentioned above are diagonalised, we are in a
position to write down all the interactions in terms of the physical
fields in the spin-${\frac1 2}$ and spin-0 sectors. We emphasize that it is
the couplings of these ${\it physical}$ fields that are going to be ultimately
related to experimental observables. Hence any phenomenological constraint
that is relevant should basically apply to them. 

Now, the physical scalar states that are dominantly charged sleptons or
sneutrinos have Higgs components in them. Similarly, there are
bound to be some gaugino (or Higgsino) admixtures in the states which are
mostly $\t$ or ${\n}_{\t}$. Consequently, the Higgs and gaugino
interactions of leptons (in the current eigenstate basis) give rise to 
trilinear interaction terms involving dominantly leptonic (or sleptonic)
fields. Similar interactions of the quarks can also give rise to L-violating
interactions. Thus we notice that starting from the bilinear interaction 
$L{H_2}$, trilinear couplings of physical states very similar to 
those conventionally parametrized by $\l$ and $\l'$ automatically emerge.    

In the interactions presented below, we have designated by $e^i,
{\n}^i ({\tilde e}^i, {\tilde {\n}}^i)$ the fermion (scalar) mass
eigenstates which are dominantly leptons (sleptons) of the $i~{th}$ generation.
The scalar (pseudoscalar) dominated by the real (imaginary) part of
${{\tilde {\n}}_L}^i$ is described as ${{\tilde {\n}}_{L1}}^i 
({{\tilde {\n}}_{L2}}^i)$. Thus we end up with trilinear 
terms in the Lagrangian,given by
\bea
{\cal L}_{tr} = {\cal L}_1 + {\cal L}_2
\eea
with
\bea
{\cal L}_1 = {\rho}_{i3i}{{\tilde e}_L}^{*i}{{\bar e}_R}^{3c}{{\n}_L}^{i} +
{{\rho}'}_{333}{{\tilde e}_L}^{3}{{\bar e}_R}^{3}{{\n}_L}^{3} +
{\omega}_{i3i}{{\tilde {\n}}_{L1}}^{i}{{\bar e}_R}^{3}{e_L}^{i}\cr + 
{{\omega}'}_{i3i}{{\tilde {\n}}_{L2}}^{i}{{\bar e}_R}^{3}{e_L}^{i} +   
{\eta}_{i3i}{{\tilde e}_L}^{*i}{{\bar {\n}}_{L}}^{3c}{e_L}^i + 
{\xi}_{i3i}{{\tilde {\n}}_{L}}^{*i}{{\bar {\n}}_{L}}^{3c}{{\n}_{L}}^{i}\cr + 
{\zeta}_{ii3}{{\tilde e}_{R}}^{i}{{\bar e}_{R}}^{i}{{\n}_{L}}^3 +
{{\zeta}'}_{333}{{\tilde e}_{R}}^{*3}{{\bar e}_{R}}^{3c}{{\n}_{L}}^3 +
{\delta}_{333}{{\tilde e}_{R}}^{*3}
{{\bar {\n}}_{L}}^{3c}{e_L}^3 + {\rm H.c.},
\eea
\bea
{\cal L}_2 = {\Omega}_{3ii}{{\bar e}_R}^{3c}{{u}_L}^{i}{{\tilde d}_L}^{*i}
+ {{\Omega}'}_{3ii}{{\bar e}_R}^{3}{{d}_L}^{i}{{\tilde u}_L}^{*i} +
{\L}_{i3i}{{\bar u}_{L}}^{i}{{\n}_{L}}^{3c}{{\tilde u}_{L}}^{i}\cr +
{{\L}'}_{i3i}{{\bar d}_{L}}^{i}{{\n}_{L}}^{3c}{{\tilde d}_{L}}^{i} +
{{\L}''}_{i3i}{{\bar u}_{R}}^{i}{{\n}_{L}}^{3}{{\tilde u}_{R}}^{i} +
{{\L}'''}_{i3i}{{\bar d}_{R}}^{i}{{\n}_{L}}^{3}{{\tilde d}_{R}}^{i}\cr +
{\Psi}_{ij3}{{\bar d}_{R}}^{i}{{u}_{L}}^{j}{{\tilde e}_{L}}^{3} +
{{\Psi}'}_{ij3}{{\bar d}_{R}}^{i}{{u}_{L}}^{j}{{\tilde e}_{R}}^{3} +
{{\Psi}''}_{3ij}{{\bar e}_{R}}^{3c}{{u}_{L}}^{i}{{\tilde d}_{R}}^{*j}\cr +
{{\Psi}'''}_{3ij}{{\bar {\n}}_{L}}^{3c}{{d}_{L}}^{i}{{\tilde d}_{R}}^{*j} +
{\Delta}_{i3j}{{\bar d}_{R}}^{i}{{e}_{L}}^{3}{{\tilde u}_{L}}^{j} +
{{\Delta}'}_{i3j}{{\bar d}_{R}}^{i}{{\n}_{L}}^{3}{{\tilde d}_{L}}^{j}\cr +
{{\Delta}''}_{ij3}{{\bar d}_{R}}^{i}{{d}_{L}}^{j}{{\tilde {\n}}_{L1}}^{3} +
{{\Delta}'''}_{ij3}{{\bar d}_{R}}^{i}{{d}_{L}}^{j}{{\tilde {\n}}_{L2}}^{3} +
{\Sigma}_{3ij}{{\bar e}_{R}}^{3}{{d}_{L}}^{i}{{\tilde u}_{R}}^{*j}\cr +
{{\Sigma}'}_{3ij}{{\bar {\n}}_{L}}^{3c}{{u}_{L}}^{i}{{\tilde u}_{R}}^{*j} +
{{\Sigma}''}_{ij3}{{\bar u}_{R}}^{i}{{d}_{L}}^{j}{{\tilde e}_{L}}^{*3} +
{{\Sigma}'''}_{ij3}{{\bar u}_{R}}^{i}{{d}_{L}}^{j}{{\tilde e}_{R}}^{*3}\cr +
{\chi}_{ij3}{{\bar u}_{R}}^{i}{{u}_{L}}^{j}{{\tilde {\n}}_{L1}}^{3} +
{{\chi}'}_{ij3}{{\bar u}_{R}}^{i}{{u}_{L}}^{j}{{\tilde {\n}}_{L2}}^{3} +
{{\chi}''}_{i3j}{{\bar u}_{R}}^{i}{{e}_{L}}^{3c}{{\tilde d}_{L}}^{j}\cr +
{{\chi}'''}_{i3j}{{\bar u}_{R}}^{i}{{\n}_{L}}^{3}{{\tilde u}_{L}}^{j} 
+ {\rm H.c.}
\eea
where the notation used for the quark and squark fields is obvious. The 
detailed expressions for the  different couplings in terms of the elements
of the mixing matrices will be found in the Appendix. Wherever the index
3 has been kept fixed in the couplings, it is because
only the third generation of leptons mixes with Higgs in this picture.

It is instructive to compare the above Lagrangian with that obtained from 
${\l}$- and ${\l}'$-type trilinear terms in the superpotential. In the
notation of reference \cite{bhg}, such interactions are 
\bea
{\cal L}_{\l} = {\l}_{ijk}[{{\tilde {\n}}_{L}}^{i}{{\bar e}_R}^{k}{e_L}^{j} +
{{\tilde e}_L}^{j}{{\bar e}_R}^{k}{{\n}_L}^{i} +
{{\tilde e}_R}^{k*}{{\bar {\n}}_{L}}^{ci}{{e}_{L}}^{j} - (i{\longleftrightarrow}
j)] + {\rm H.c.},
\eea   
\bea
{\cal L}_{{\l}'} = {{\l}'}_{ijk}[{{\tilde {\n}}_{L}}^{i}{{\bar d}_R}^{k}
{d_L}^{j} +
{{\tilde d}_L}^{j}{{\bar d}_R}^{k}{{\n}_L}^{i} +
{{\tilde d}_R}^{k*}{{\bar {\n}}_{L}}^{ci}{{d}_{L}}^{j}
-{{\tilde e}_{L}}^{i}{{\bar d}_R}^{k}
{u_L}^{j} -
{{\tilde u}_L}^{j}{{\bar d}_R}^{k}{e_L}^{i} -
{{\tilde d}_R}^{k*}{{\bar e}_{L}}^{ci}{{u}_{L}}^{j}]
+ {\rm H.c.}
\eea
First of all, all the terms in equation (23) can be generated in ${\cal L}_1$
if one allows for mixing of the three leptonic generations and also Yukawa
coupling of the first two generations. The fact that we have neglected
both of the above features is responsible for the absence of coefficients
in equation (21) with all three indices different. On the other hand, 
${\cal L}_1$ can contain coefficients with generation indices \{iii\} (in
our case, \{333\} only because of reasons stated above). Such terms are 
forbidden in equation (23) by gauge invariance of the superpotential, unless
there is mixing among the lepton generations. 

Next, we note that the terms in ${\cal L}_1$ proportional to $\rho$, $\eta$,
$\xi$, $\zeta$ and ${\zeta}'$ do not arise in (23). The $\rho$-term
owes its structure to the gaugino and Higgsino couplings in the MSSM part
of the Lagrangian. The four remaining terms also could not be allowed in (23)
because, again, $SU(2)$ invariance of the superpotential would forbid
terms with either three left chiral fields or one left and two right
chiral fields. A particularly interesting consequence of this is the presence
of trilinear interaction involving a sneutrino and two neutrino physical
fields. This lends considerable additional phenomenology to the scenario under
study here. For example, a sneutrino can decay into two neutrinos here 
\cite{anj}. Also, as we shall see below, it entails the possibility of 
invisible decays of the lightest neutralino.

On comparing equations (22) and (24) we find that all the ${\l}'$-type
terms are generated in ${\cal L}_2$ as well. In addition, there are
several more terms which are prevented in (24) in order to prevent weak
isospin and hypercharge violation in the superpotential.

The other novel consequences of the $L{H_2}$ term are the flavor-changing
couplings of the $W$ and the $Z$. Although it has been sometimes claimed 
in the literature \cite{zw} to be a signature of spontaneous $R$-parity
violation, in practice it follows just from the bilinear terms of the type
discussed by us. Diagonalisation of the mass matrices 
$M_{{\tilde {\chi}}^{\pm}}$ and $M_{{\tilde {\chi}}^0}$ immediately imply
that now there can be a tree-level interaction involving a ${\t} ({\n}_{\t})$
dominated physical state, a neutralino (chargino)-dominated state and a $W$.
Similarly, the fact that the neutralinos (charginos) and the ${\n}_{\t}(\t)$
differ in $T_3$ and $Y$ implies that their $Z$-couplings can now be 
non-diagonal. The interactions are given by
\bea
{\cal L}_{{W^-}{{\tilde \chi}^+}{{\tilde \chi}^0}} = g {{W_{\m}}^-}
{\bar {{\tilde \chi}^0}}_{i}{\g}^{\m}[~O_{ij}^L{P_L} + O_{ij}^R{P_R}~]
{{\tilde \chi}_{j}}^{+} + {\rm H.c.},
\eea
\bea
{\cal L}_{{\bar {\tilde \chi}}{\tilde \chi}{Z}} = \frac g {{\rm cos}
{{\theta}_{W}}}[~{\bar {\tilde \chi}}_{i}{\g}^{\m}({O'}_{ij}^L{P_L}+ 
{O'}_{ij}^R{P_R}){\tilde \chi}_{j}~]{Z_{\m}} + {\rm H.c.},
\eea
\bea
{\cal L}_{{\bar {\tilde \chi}^0}{{\tilde \chi}^0}{Z}} = \frac g {{\rm cos}
{{\theta}_{W}}}[~{\frac 1 2}{{\bar {\tilde \chi}}^{0}}_{i}{\g}^{\m}
({O''}_{ij}^L{P_L} + {O''}_{ij}^R{P_R}){{\tilde \chi}^{0}}_{j}~]
{Z_{\m}} + {\rm H.c.}
\eea   
where the detailed forms of the matrices $O$, $O'$ and $O''$ are
relegated to the Appendix.

We end this section by re-iterating that the bilinear interaction $L{H_2}$
is sufficient to generate all the $\l$- and ${\l}'$-type terms involving
${\it {physical}}$ fields. They also give rise to other trilinear interaction
terms which are otherwise disallowed. Furthermore, the postulate that only
the third generation is involved in $L$-violating mixing (partially justified 
by the observed mass hierarchy) suggests that flavor changing trilinear
terms should be smaller in magnitude.

\section{Numerical Results}

\noindent In order to find out the allowed region in the parameter space,
one has to take a number of constraints into account. First, we note that $\e$
and $v_3$ are the only parameters outside MSSM that enter into the chargino 
and neutralino mass matrices. The strongest constraint on them follows
from the fact that the $\t$-mass has been experimentally measured 
\cite{pdg}. 
Therefore, for any combination of the MSSM parameters $(m_{\tilde g}, \m, 
{{\rm tan}{\b}})$, the lowest eigenvalue of $m_{{\tilde \chi}^{\pm}}$ should
agree with $m_{\t}$ for any combination of $\e$ and $v_3$. Also, $\n_{\t}$
has a laboratory upper limit of 24 MeV on its mass. These two restrictions,
taken together, constrain the $\e-v_3$ space in a severe manner.

Figures $(1-4)$ show the allowed areas of the $\e-v_3$ parameter space for
several combinations of the MSSM parameters. Here, in addition to the 
constraint $m_{\n_{\t}}~<~$24 MeV the lowest eigenvalue of the 
$m_{{\tilde \chi}^{\pm}}$ has been allowed at most a 3$\sigma$-deviation
from the measured central value of $m_{\t}$. However, there is an extra
parameter here to play with, namely, the third diagonal entry of 
$m_{{\tilde \chi}^{\pm}}$. In the figures presented here, we have fixed
this term at the central experimental value of $m_{\t}$, viz. 1.777 GeV.
The allowed area slightly increases on varying this mass parameter, but it
is not permissible to drift too much from the value used here. It is easy
to check that our allowed region is consistent at 95 $\%$ 
confidence level with the restrictions imposed by the global fit of LEP 
data and low energy experiments on the mixing of the $\t$ and the 
${\n}_{\t}$ with exotic fermions \cite{nardi}.
In any case, we find that there is no substantial allowed region 
with $\vert\e\vert$ and $v_3$ larger than about 20 and 5 GeV respectively. 
Sometimes there are extremely narrow allowed bands with one of them of a 
considerably higher value. Such ``fine-tuned" areas are not used in our 
subsequent calculations. 

The next set of constraints arise from the scalar sector where all of the
four conditions mentioned in the previous section have to be fulfilled.
The scalar potential introduces several new parameters: $m_1, m_2, m_0$ 
(the slepton/sneutrino mass assuming a degeneracy), $A, B_1$ and $B_2$. 
Of these, the minimisation conditions imply that only three are independent. 
We have chosen $A$, $B_1$ and $B_2$ to be the three independent parameters. 
Furthermore, we set $A$ equal to zero to simplify our analysis. 
Now, taking $\e$ and $v_3$ from the allowed regions described above, 
one gets restricted in the choice of $B_1$ and $B_2$. A definite requirement 
in this respect is that $B_1(B_2)$ should have a sign opposite to that of 
$\m(\e)$. 

Having thus been guided to the allowed region in the entire parameter space, 
we can now compute all the $R$-parity violating couplings in terms of them. 
We have neglected all CP-violating phases.
The values of these for some sample values of the SUSY parameters are
shown in Tables 1 and 2. The numbers indicate the maximum values that the 
respective couplings can have. Most of the couplings are seen to be on the
order of $10^{-3}$ or less, excepting a few on the order of $10^{-2}$ or even
$10^{-1}$. It is noticeable that a higher value of $\e$ often raises the 
couplings. The cases where this does not happen can be ascribed to enhanced
cancellations among the different terms that comprise a particular coupling.
In particular, an enhancement in some of the terms occurs when the slepton
mass $m_0$ is close to one of the Higgs masses, which causes a large 
slepton(sneutrino)- Higgs mixing. Wherever such mixing terms dominate in
any interaction strength, the corresponding strength is large.

In general, the $R$-violating interactions that we obtain here after satisfying
all requisite constraints are considerably smaller than the bounds on the 
analogous $\l$- and $\l'$-type terms derived in the literature from
existing experimental data. The latter includes limits from a wide variety
of phenomena, from low-energy weak processes to results from the Large
Electron Positron (LEP) collider. This suggests that if indeed bilinear 
interactions are the real sources of the nonconservation of $R$-parity, then
our experimental precision requires considerable improvement before such
interactions can be probed. 

Finally, let us turn to some processes that can be looked upon as the
typical consequences of bilinear $R$-violating terms. Of course, the
lightest neutralino ${\tilde \chi}^0$ (the LSP in MSSM) is bound to be 
unstable. When its mass is less than that of the standard gauge bosons, it
can only have three-body decays. The final states for such decays are the same
whether $R$-parity is violated originally through bilinear or trilinear 
interactions. 

However, if $m_{{\tilde \chi}^0}$ is larger than $m_Z$, $m_W$, then the
bilinear terms in our scenario open up two-body decay channels which are not
otherwise possible. These are the channels ${\tilde \chi}^{0}
\longrightarrow {\t}W$ and ${\tilde \chi}^{0} \longrightarrow {{\n}_{\t}}Z$,
controlled by $O_{L(R)}$, ${O'}_{L(R)}$ and ${O''}_{L(R)}$ of 
equations 25-27. In Table 3 we list their values for the same set of input
parameters as in Tables 1 and 2.

Figures 5 and 6 contain some plots for the dominant branching ratios,
assuming that ${\tilde \chi}^{0} \longrightarrow {\t}W$ and 
${\tilde \chi}^{0} \longrightarrow {{\n}_{\t}}Z$
are the only available channels. The former mode is found to dominate in
figure 5, while the latter takes over in figure 6. This is because there 
are essentially two main components in each of the 
${\tilde \chi}^{0}{\t}W$ and ${\tilde \chi}^{0}{{\n}_{\t}}Z$ 
interactions. One of these comes from the neutrino-tau(neutrino)-W(Z) gauge
couplings, and the other, from the Higgsino-Higgsino-W(Z) coupling. In the
area of the parameter space shown in figure 5, it is found that while the 
two components add up in the former process, they interfere destructively in 
the latter, causing a large cancellation. Exactly the opposite thing happens
in figure 6, where, in particular, $\vert \m \vert$, $\vert B_1 \vert$ and
$\vert B_2 \vert$ are large. This indicates that pair-produced neutralinos 
are expected to give rise to signals of the form ${\t}{\t}WW$ and ZZ + 
${\not p}_T$.  $R$-parity violation through bilinear interaction 
terms is quite characteristically reflected through such signals. 

\section{Summary and Conclusions}

\noindent We have studied the effects of an $R$-parity violating bilinear term
$L_3{H_2}$ in the superpotential. We find that this term, together with a
sneutrino vev, leads to trilinear couplings
involving dominantly leptonic and sleptonic (as also two quarks/squarks and
one lepton/slepton) physical fields. We emphasize that it is these terms 
involving ${\it physical}$ fields which are of phenomenological significance.
The interactions thus generated include the $\l$- and ${\l}'$-type ones
which follow from trilinear $R$-parity violating terms in the superpotential.
In addition, we obtain several terms that are not permitted in the other case.
The most noteworthy among them is the one involving a sneutrino and two 
neutrinos. Also, there arise off-diagonal interactions of charginos and
neutralinos with $\t$ and ${\n}_{\t}$ coupled to a $W$ or a $Z$. Such 
interactions are the characteristic features of bilinear $R$-violation. 
The parameter space of such a scenario can be best limited by restricting 
the lowest eigenvalues of the chargino and neutralino mass matrices. Further
constraints follow from requirements of electroweak symmetry breaking in the 
scalar sector. The trilinear couplings thus generated mostly turn out to be 
small compared to their current phenomenological limits. Thus if bilinear 
terms are the sole sources of $R$-parity violation, then the restrictions 
imposed by the lepton and neutrino masses are still more stringent than any 
other experimental bound. And finally, we have discussed some phenomenological 
consequences of such a scenario. In particular, we show that if the
lightest neutralino is heavier than the weak gauge bosons, then its dominant
decay occurs in the channels ${\tilde \chi}^{0} \longrightarrow {\t}W$ and
${\tilde \chi}^{0} \longrightarrow {{\n}_{\t}}Z$ giving rise to rather 
characteristic signals.
\bigskip

\bigskip
\noindent {\large {\bf Acknowledgement}}

\vspace{.2in}
\noindent We thank P. K. Mohanty, A. Rastogi, Suresh Rao and A. Ghosal for 
computational help and A. Kundu  and A. Datta for helpful discussions.

\newpage
\appendix
\renewcommand{\theequation}{A.{\arabic{equation}}}
\setcounter{equation}{0}
\bigskip

\bigskip
\noindent {\large {\bf Appendix~A}}

\noindent Here we present the full forms of the various couplings in equations (21) 
and (22). In obtaining the trilinear interactions, the mixing among
leptons and quarks in gaugino couplings have been neglected. Using the 
notation already established in the text,
\bea
{\rho}_{131} = {\rho}_{232} = -g {U^*}_{31}
\eea
\bea
{\rho}_{333} = -g {U^*}_{31} + f {U^*}_{32} {N^*}_{55} {C_{43}}
\eea
\bea
{{\rho}'}_{333} = f {V^*}_{33} {N^*}_{55} {C_{23}} - 
f {V^*}_{33} {N^*}_{53} {C_{33}}
\eea
\bea
{\omega}_{131} = {\omega}_{232} = - g {V^*}_{31}
\eea
\bea
{\omega}_{333} = - g {V^*}_{31} + \frac 1 {\sqrt 2} f {V^*}_{33} {U^*}_{32}
{S_{33}} - \frac 1 {\sqrt 2} f {V^*}_{33} {U^*}_{33} {S_{13}}
\eea
\bea
{\omega'}_{131} = {\omega'}_{232} = g {V^*}_{31}
\eea
\bea
{\omega'}_{333} = g {V^*}_{31} + \frac 1 {\sqrt 2} f {V^*}_{33} {U^*}_{32}
{P_{33}} - \frac 1 {\sqrt 2} f {V^*}_{33} {U^*}_{33} {P_{13}}
\eea
\bea
{\eta}_{131} = {\sqrt 2}~e {N^*}_{51} - \frac {{\sqrt 2}~g } {{\rm cos}
{{\theta}_W}} (-\frac 1 2 + {{\rm sin}^2}{{\theta}_W}) {N^*}_{52}
\eea
\bea 
{\eta}_{232} = {\sqrt 2}~e {N^*}_{51} - \frac {{\sqrt 2}~g } {{\rm cos}
{{\theta}_W}} (-\frac 1 2 + {{\rm sin}^2}{{\theta}_W}) {N^*}_{52}
\eea
\bea
{\eta}_{333} = {\sqrt 2}~e {N^*}_{51} - \frac {{\sqrt 2}~g } {{\rm cos}
{{\theta}_W}} (-\frac 1 2 + {{\rm sin}^2}{{\theta}_W}) {N^*}_{52} -
f {N^*}_{53} {U^*}_{33} {C_{43}}
\eea
\bea
{\xi}_{131} = - \frac {{\sqrt 2}~g } {2 {{\rm cos}
{{\theta}_W}}} {N^*}_{52}
\eea
\bea
{\xi}_{232} = - \frac {{\sqrt 2}~g } {2 {{\rm cos}
{{\theta}_W}}} {N^*}_{52}
\eea
\bea
{\xi}_{333} = - \frac {{\sqrt 2}~g } {2 {{\rm cos}
{{\theta}_W}}} {N^*}_{52}
\eea 
\bea
{\zeta}_{113} = -{\sqrt 2}~e {N^*}_{51} + \frac {{\sqrt 2}~g 
{{{\rm sin}^2}{{\theta}_W}}} {{\rm cos}
{{\theta}_W}} {N^*}_{52}
\eea
\bea
{\zeta}_{223} = -{\sqrt 2}~e {N^*}_{51} + \frac {{\sqrt 2}~g 
{{{\rm sin}^2}{{\theta}_W}}} {{\rm cos}
{{\theta}_W}} {N^*}_{52}
\eea
\bea
{\zeta}_{333} = -{\sqrt 2}~e {N^*}_{51} + \frac {{\sqrt 2}~g 
{{{\rm sin}^2}{{\theta}_W}}} {{\rm cos}
{{\theta}_W}} {N^*}_{52} + f {V^*}_{33} {N^*}_{55} {C_{24}} -
f {V^*}_{33} {N^*}_{53} {C_{34}}  
\eea
\bea
{{\zeta}'}_{333} = f {U^*}_{32} {N^*}_{55} {C_{44}}
\eea
\bea
{\delta}_{333} = -f {U^*}_{33} {N^*}_{53} {C_{44}}
\eea
\bea
{\Omega}_{3ii} = - g {U^*}_{31}
\eea
\bea
{\Omega'}_{3ii} = - g {V^*}_{31}
\eea
\bea
{\L}_{i3i} = -{\sqrt 2}\{{\frac g {{\rm cos}{\theta}_{W}}}({\frac 1 2} -
{\frac 2 3}{{\rm sin}^2}{\theta}_{W}) N_{52} + {\frac 2 3} g~ {\rm sin}
{\theta}_{W} N_{51}\} 
\eea
\bea
{{\L}'}_{i3i} = {\sqrt 2}\{{\frac g {{\rm cos}{\theta}_{W}}}(-{\frac 1 2} +
{\frac 1 3}{{\rm sin}^2}{\theta}_{W}) N_{52} - {\frac 1 3} g~ {\rm sin}
{\theta}_{W} N_{51}\} 
\eea 
\bea
{{\L}''}_{i3i} = -{\sqrt 2}\{{\frac 2 3}{\frac g {{\rm cos}{\theta}_{W}}}
{{\rm sin}^2}{\theta}_{W} {N^*}_{52} - {\frac 2 3} g~ {\rm sin}
{\theta}_{W} {N^*}_{51}\} 
\eea
\bea
{{\L}'''}_{i3i} = {\sqrt 2}\{- {\frac 1 3}{\frac g {{\rm cos}{\theta}_{W}}}
{{\rm sin}^2}{\theta}_{W} {N^*}_{52} + {\frac 1 3} g~ {\rm sin}
{\theta}_{W} {N^*}_{51}\} 
\eea
\bea
{\Psi}_{ij3} = {f_1} C_{23}
\eea
\bea
{{\Psi}'}_{ij3} = {f_1} C_{24}
\eea
\bea
{{\Psi}''}_{3ij} = {f_1} {U^*}_{32}
\eea
\bea
{{\Psi}'''}_{3ij} = - {f_1} {N^*}_{53}
\eea
\bea
{\Delta}_{i3j} = {f_1} {U^*}_{32}
\eea
\bea
{{\Delta}'}_{i3j} = -{f_1} {N^*}_{53}
\eea
\bea
{{\Delta}''}_{ij3} = -{\frac 1 {\sqrt 2}}{f_1} {S_{13}}
\eea
\bea
{{\Delta}'''}_{ij3} = -{\frac 1 {\sqrt 2}}{f_1} {P_{13}}
\eea
\bea
{\Sigma}_{3ij} = {f_2} {V^*}_{32} 
\eea
\bea
{{\Sigma}'}_{3ij} = -{f_2}{N^*}_{54} 
\eea
\bea
{{\Sigma}''}_{ij3} = {f_2}{C_{13}} 
\eea
\bea
{{\Sigma}'''}_{ij3} = {f_2}{C_{14}} 
\eea
\bea
{\chi}_{ij3} = -{\frac 1 {\sqrt 2}} {f_2} {S_{23}}
\eea
\bea
{{\chi}'}_{ij3} = -{\frac 1 {\sqrt 2}} {f_2} {P_{23}}
\eea
\bea
{{\chi}''}_{i3j} = {f_2} {V^*}_{23}
\eea
\bea
{{\chi}'''}_{i3j} = - {f_2} {N^*}_{54}
\eea
$f_1$ and $f_2$ are defined above as $f_1 = h^{d}_{33} = {\frac {m_b} {v_1}}$,
$f_2 = h^{u}_{33} = {\frac {m_t} {v_2}}$. We have neglected the Yukawa 
interactions of the remaining quarks. 

\noindent Next we give detailed forms of the matrices $O$, $O'$ and 
$O''$ which appear in equations (25)-(27).
\bea
O_{ij}^L = {\frac 1 {\sqrt 2}} N_{i4} {V^*}_{j2} - {\rm cos}{{\theta}_{W}} 
N_{i2} {V^*}_{j1} - {\rm sin}{\theta}_{W} N_{i1} {V^*}_{j1} 
\eea
\bea
O_{ij}^R = -{\frac 1 {\sqrt 2}} {N^*}_{i3} {U_{j2}} - {\rm cos}{{\theta}_{W}}
{N^*}_{i2} {U_{j1}} - {\rm sin}{\theta}_{W} {N^*}_{i1} {U_{j1}}
+ {\frac 1 {\sqrt 2}} {N^*}_{i5} {U_{j3}} 
\eea
\bea
{O'_{ij}}^L = V_{i1} {V^*}_{j1} + {\frac 1 2} V_{i2} {V^*}_{j2} 
- {\delta}_{ij} {{\rm sin}^2}{\theta}_{W} + V_{i3} {V^*}_{j3} {{\rm sin}^2}
{\theta}_{W}
\eea
\bea
{O'_{ij}}^R = {U^*}_{i1} U_{j1} + {\frac 1 2} {U^*}_{i2} U_{j2} 
- {\delta}_{ij} {{\rm sin}^2}{\theta}_{W} + {U^*}_{i3} U_{j3} (-{\frac 1 2} +
{{\rm sin}^2}{\theta}_{W})
\eea  
\bea
{O''_{ij}}^L = {\frac 1 2}N_{i3} {N^*}_{j3} - {\frac 1 2} N_{i4} 
{N^*}_{j4} + {\frac 1 2} N_{i5} {N^*}_{j5}
\eea
\bea
{O''_{ij}}^R = -{\frac 1 2}{N^*}_{i3} N_{j3} + {\frac 1 2} {N^*}_{i4} 
N_{j4} = - {{O''}_{ij}}^L
\eea

\newpage
\begin{center}
{\bf Table 1}
\end{center}
\begin{center}
\begin{tabular}{|l|r|l|} \hline\hline
\multicolumn{2}{|r|}{$\e$ = 16} & {$\e$ = 2} \\ \hline
{${\rho}_{131}$} & 0.00633 & -0.00589 \\ 
{${\rho}_{232}$} & 0.00633 & -0.00589 \\
{${\rho}_{333}$} & 0.00776 & -0.00597 \\
{${{\rho}'}_{333}$} & -8.1 $\times$ ${10}^{-6}$ & -0.00014 \\
{${\omega}_{131}$} & -0.00018 & 0.00017 \\
{${\omega}_{232}$} & -0.00018 & 0.00017 \\
{${\omega}_{333}$} & -0.00027 & -0.00036 \\
{${{\omega}'}_{131}$} & 0.00018 & -0.00017 \\
{${{\omega}'}_{232}$} & 0.00018 & -0.00017 \\
{${{\omega}'}_{333}$} & 0.00020 & -0.00031 \\ \hline
\end{tabular} 
\hspace{0.0in}
\begin{tabular}{|l|r|l|} \hline\hline
\multicolumn{2}{|r|}{$\e$ = 16} & {$\e$ = 2} \\ \hline
{${\eta}_{131}$} & -0.00137 & 0.00128 \\
{${\eta}_{232}$} & -0.00137 & 0.00128 \\
{${\eta}_{333}$} & -0.00279 & 0.00137 \\
{${\xi}_{131}$} & 0.00422 & -0.00393 \\
{${\xi}_{232}$} & 0.00422 & -0.00393 \\
{${\xi}_{333}$} & 0.00422 & -0.00393 \\
{${\zeta}_{113}$} & -0.00413 & 0.00384 \\
{${\zeta}_{223}$} & -0.00413 & 0.00384 \\
{${\zeta}_{333}$} & -0.00413 & 0.00399 \\
{${{\zeta}'}_{333}$} & 0.00128 & -0.00009 \\
{${\delta}_{333}$} & -0.00127 & 0.00010 \\ \hline
\end{tabular}
\end{center}
\bigskip

\bigskip
\vspace{.2in}

\noindent Table 1:

\noindent Sample values of the couplings in equation 21, for
two values of $\e$, with $v_3$ = 3.4, $\m$ = 200, $m_{\tilde g}$ = 750,
$B_1$ = -180, $B_2$ = -160, ${\rm tan} {\b}$ = 2. All the mass parameters
are expressed in GeV.

\newpage
\begin{center}
{\bf Table 2}
\end{center}
\begin{center}
\begin{tabular}{|l|r|l|} \hline\hline
\multicolumn{2}{|r|}{$\e$ = 16} & {$\e$ = 2} \\ \hline
{${\Omega}_{3ii}$} & 0.00633 & -0.00589 \\ 
{${{\Omega}'}_{3ii}$} & -0.00018 & 0.00017 \\
{${\L}_{i3i}$} & 0.00275 & -0.00256 \\
{${{\L}'}_{i3i}$} & 0.00584 & -0.00543 \\
{${{\L}''}_{i3i}$} & 0.00235 & -0.00219 \\
{${{\L}'''}_{i3i}$} & 0.00117 & -0.00109 \\
{${\psi}_{ij3}$} & 0.00308 & -0.00059 \\
{${{\psi}'}_{ij3}$} & -0.00344 & 0.00057 \\
{${{\psi}''}_{3ij}$} & -0.00467 & -0.00029 \\
{${{\psi}'''}_{3ij}$} & 0.00464 & 0.00033 \\
{${\Delta}_{i3j}$} & -0.00467 & -0.00029 \\ \hline
\end{tabular} 
\hspace{0.0in}
\begin{tabular}{|l|r|l|} \hline\hline
\multicolumn{2}{|r|}{$\e$ = 16} & {$\e$ = 2} \\ \hline
{${{\Delta}'}_{i3j}$} & 0.00464 & 0.00033 \\
{${{\Delta}''}_{ij3}$} & 0.00309 & -0.00151 \\
{${{\Delta}'''}_{ij3}$} & 0.00335 & -0.00054 \\
{${\Sigma}_{3ij}$} & 0.00048 & -0.00045 \\
{${{\Sigma}'}_{3ij}$} & -0.00238 & 0.00223 \\
{${{\Sigma}''}_{ij3}$} & 0.01521 & 0.01113 \\
{${{\Sigma}'''}_{ij3}$} & -0.01649 & -0.01182 \\
{${\chi}_{ij3}$} & -0.01785 & -0.04348 \\
{${{\chi}'}_{ij3}$} & 0.01674 & 0.01170 \\
{${{\chi}''}_{i3j}$} & 0.00048 & -0.00045 \\
{${{\chi}'''}_{i3j}$} & -0.00238 & 0.00223 \\ \hline
\end{tabular}
\end{center}
\bigskip

\bigskip
\vspace{.2in}

\noindent Table 2:

\noindent Sample values of the couplings in equation 22, with the same 
input parameters as in Table 1.
\newpage
\begin{center}
{\bf Table 3}
\end{center}
\begin{center}
\begin{tabular}{|l|r|l|} \hline\hline
\multicolumn{2}{|r|}{$\e$ = 16} & {$\e$ = 2} \\ \hline
$O_{51}^L$ & 0.00412 & -0.00385 \\
$O_{51}^R$ & 0.06329 & -0.01545 \\
$O_{52}^L$ & -0.00640 & 0.00595 \\
$O_{52}^R$ & 0.10156 & 0.00201 \\
$O_{43}^L$ & 0.00016 & -0.00015 \\
$O_{43}^R$ & -0.05506 & 0.00550 \\
${O'}_{32}^L$ & -0.00023 & 0.00022 \\
${O'}_{32}^R$ & -0.05813 & -0.00390 \\
${O'}_{31}^L$ & 0.00006 & -0.00006 \\
${O'}_{31}^R$ & -0.03097 & 0.01152 \\
${O''}_{45}^L$ & -0.00501 & 0.00467 \\
${O''}_{45}^R$ & -0.01581 & -0.00146 \\ \hline
\end{tabular}
\end{center}
\noindent Table 3:

\noindent Sample values of the couplings in equations 25-27, with the 
same input parameters as in Table 1.
\newpage
\centerline {\large {\bf Figure Captions}}

\hspace*{\fill}

\hspace*{\fill}

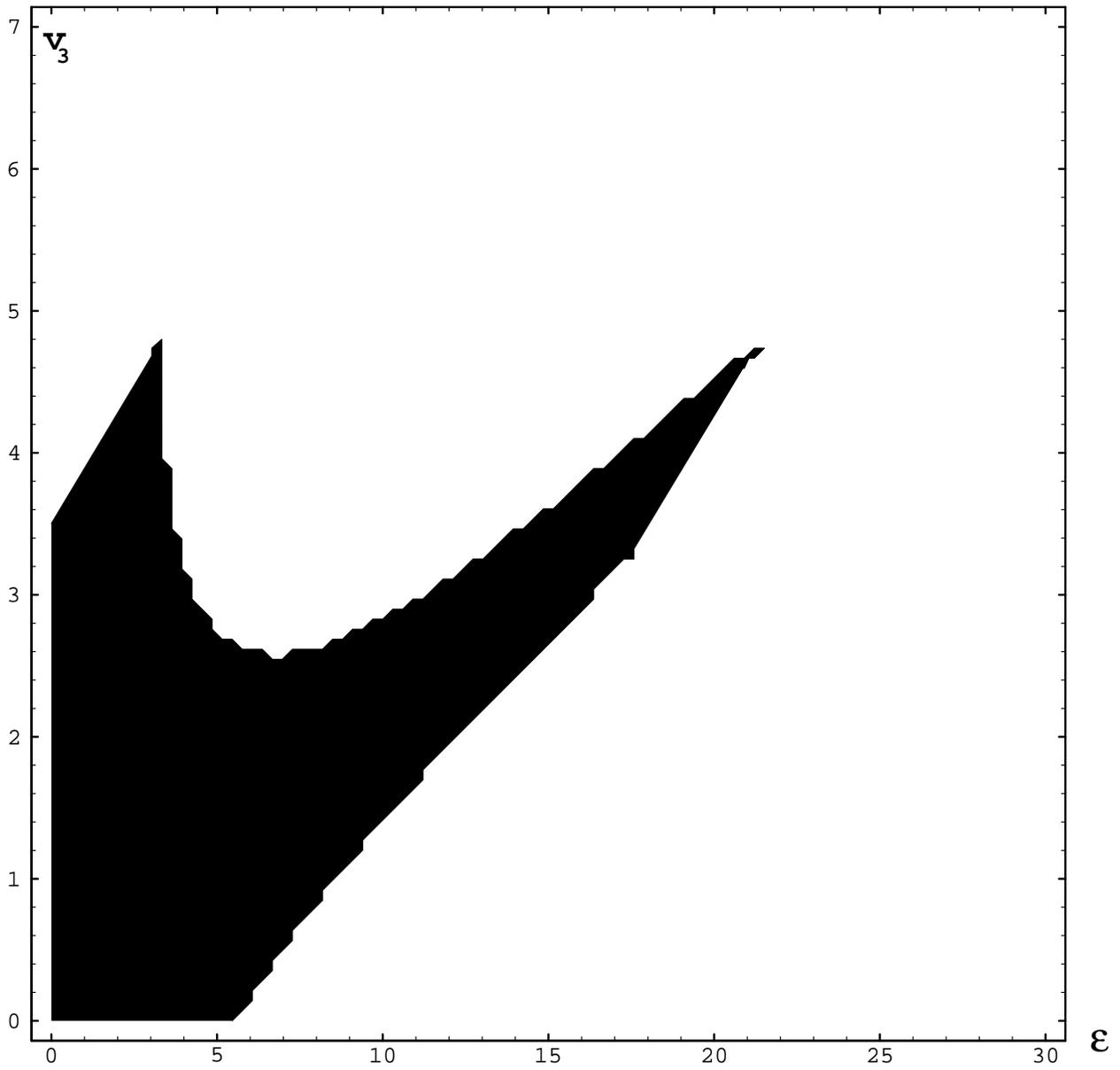
\noindent Figure 1: 

\noindent
The allowed region(dark) in the $\e$-$v_3$ parameter space,
with $B_1$ = -180, $B_2$ = -160, $\m$ = 200, ${\rm tan}{\b}$ = 2,
$m_{\tilde g}$ = 750. All mass parameters are expressed in GeV.
\vskip .25in

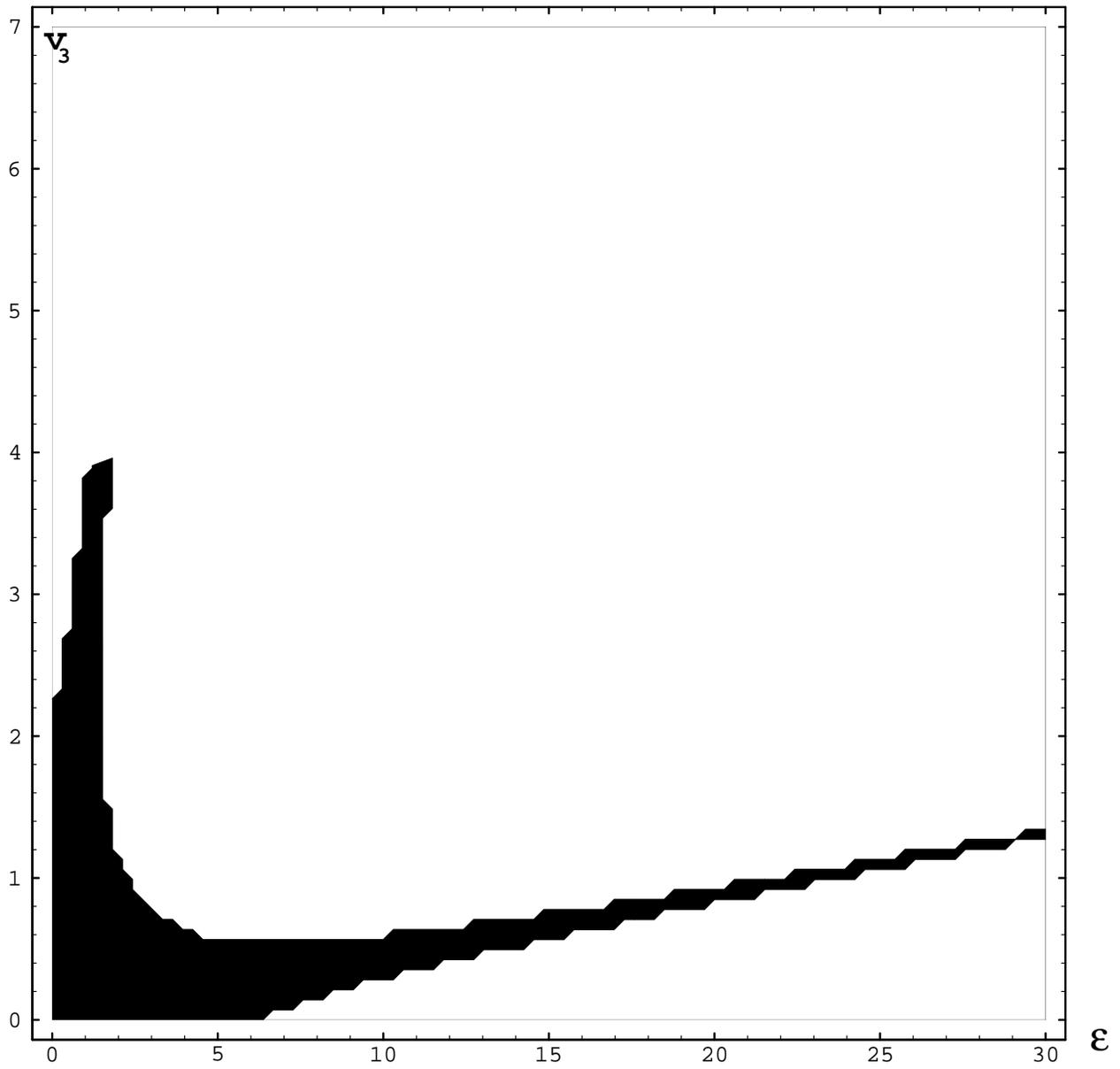
\noindent Figure 2: 

\noindent

Same as in Figure 1, with $B_1$ = -160, $B_2$ = -170, $\m$ = 200, 
${\rm tan}{\b}$ = 10, $m_{\tilde g}$ = 750.
\vskip .25in

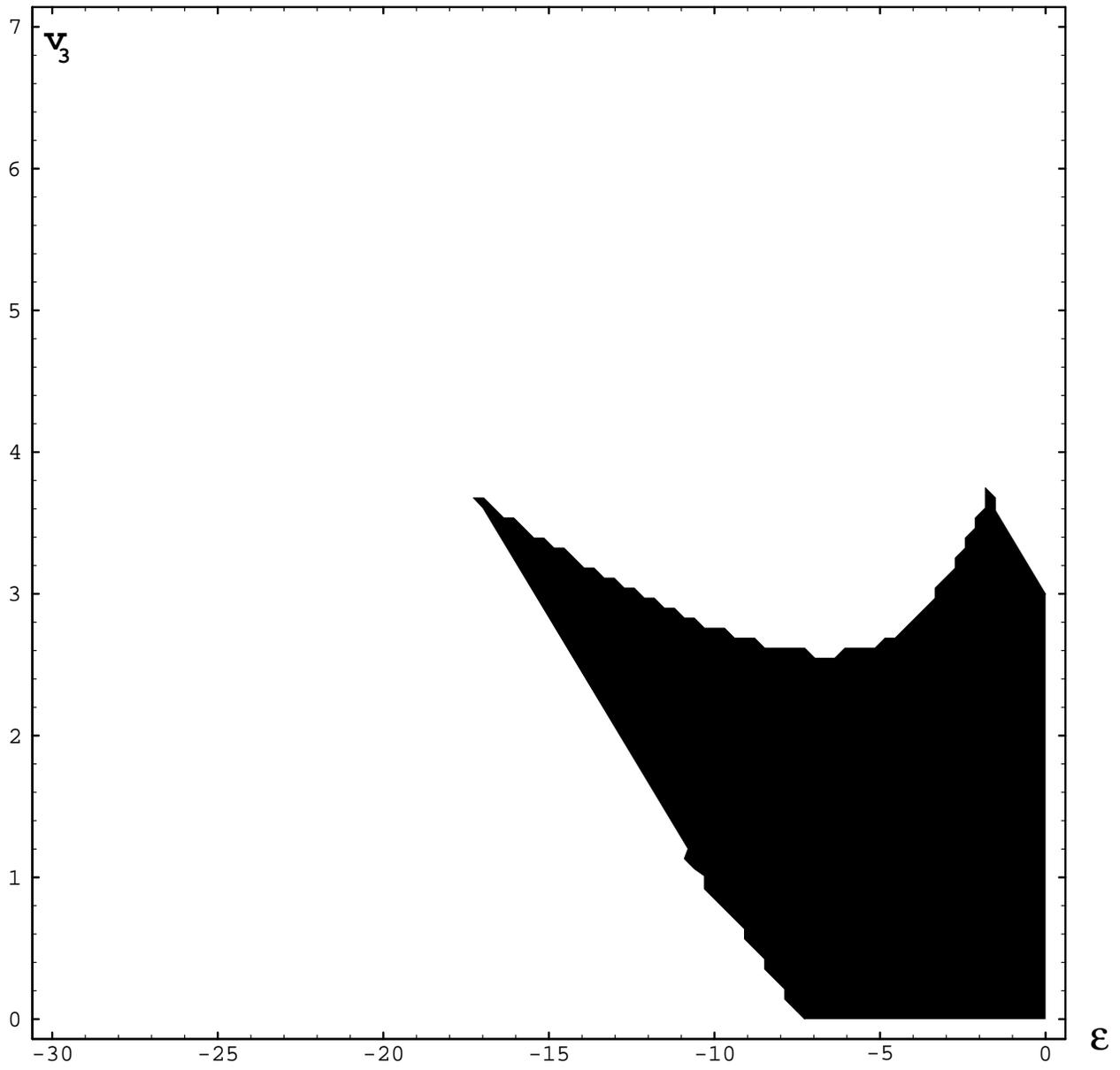
\noindent Figure 3: 

\noindent

Same as in Figure 1, with $B_1$ = 150, $B_2$ = 170, $\m$ = -200, 
${\rm tan}{\b}$ = 2, $m_{\tilde g}$ = 300.
\vskip .25in

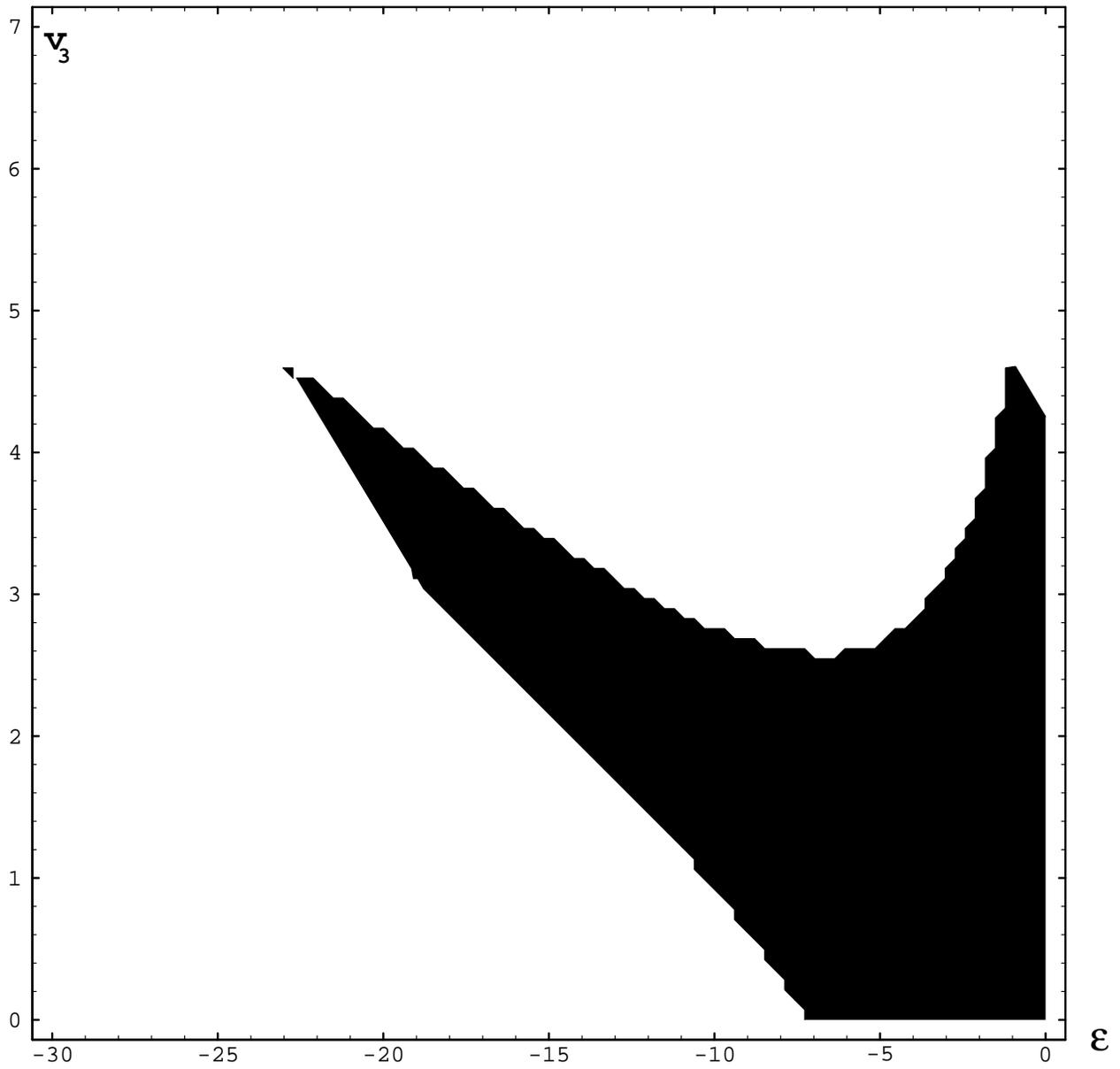
\noindent Figure 4: 

\noindent

Same as in Figure 1, with $B_1$ = 150, $B_2$ = 200, $\m$ = -200, 
${\rm tan}{\b}$ = 2, $m_{\tilde g}$ = 750.
\vskip .25in

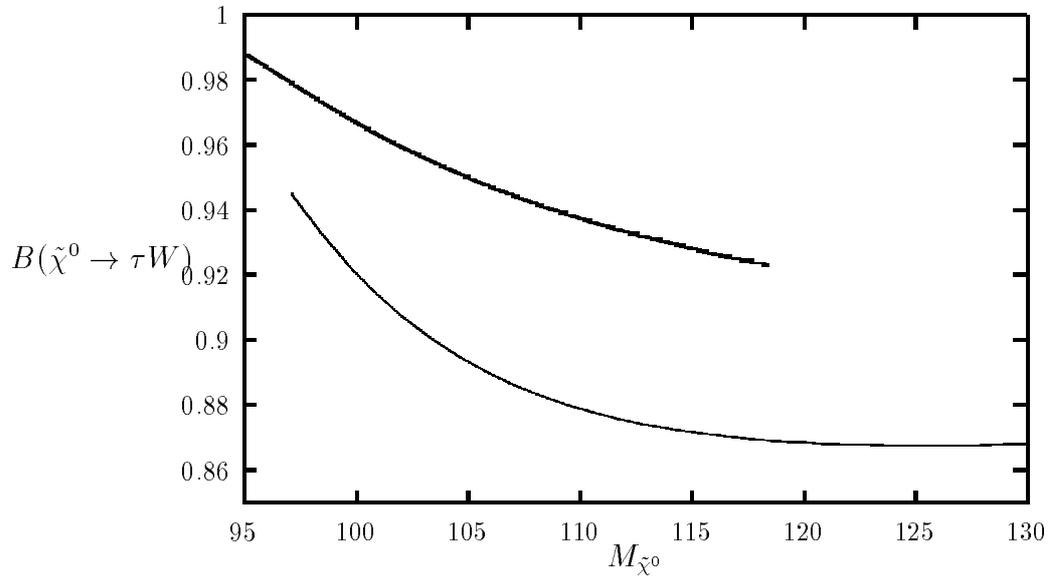
\noindent Figure 5:

\noindent 
$B({\tilde \chi}^0 \longrightarrow {\t}{W})$ plotted against
the lightest neutralino mass (in GeV). The bold (thin) line corresponds
to ${\rm tan}{\b}$ = 2, $B_1$ = -180, $B_2$ = -160, $\e$ = 5.0, $v_3$ = 2.0
$\m$ = 200 $({\rm tan}{\b}$ = 10, $B_1$ = -160, $B_2$ = -170, $\e$ = 1.0,
$v_3$ = 0.4, $\m$ = 200). All mass parameters are expressed in GeV. 
\vskip .25in

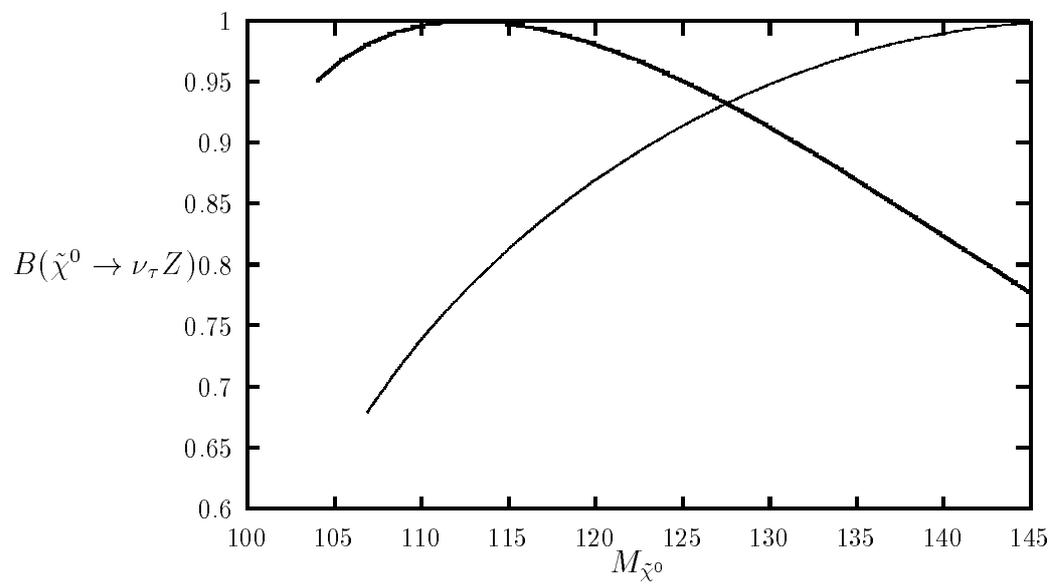
\noindent Figure 6:

\noindent
$B({\tilde \chi}^0 \longrightarrow {\n}_{\t}{Z})$ plotted against
the lightest neutralino mass (in GeV). The bold (thin) line corresponds
to ${\rm tan}{\b}$ = 2, $B_1$ = -350, $B_2$ = -330, $\e$ = 10.0, $v_3$ = 3.0
$\m$ = 500 $({\rm tan}{\b}$ = 10, $B_1$ = -280, $B_2$ = -290, $\e$ = 4.0,
$v_3$ = 1.0, $\m$ = 500).

\end{document}